\documentclass{ws-mpla}
\usepackage[super]{cite}
\usepackage{graphicx}
\begin{document}

\markboth{Plamen P Fiziev}
{Compact static stars in minimal dilatonic gravity}

\catchline{}{}{}{}{}

\title{Compact static stars in minimal dilatonic gravity}

\author{\footnotesize PLAMEN P.  FIZIEV}

\address{Sofia University Foundation of Theoretical and Computational Physics and Astrophysics, Boulevard
5 James Bourchier\\
 Sofia 1164, Bulgaria\\
fizev@phys.uni-sofia.bg\\
and\\
Bogoliubov Laboratory of Theoretical Physics, JINR, Dubna\\
141980 Moscow Region, Rusia\\
fizev@theor.jinr.ru}

\maketitle

\pub{Received (Day Month Year)}{Revised (Day Month Year)}

\begin{abstract}
In the version \cite{Fiziev14} of this paper we presented for the first time the basic equations and relations
 for relativistic static spherically symmetric  stars (SSSS)
 in the  model of minimal dilatonic  gravity  (MDG).
 This model is {\em locally} equivalent to the f(R) theory of gravity and gives an alternative
 description of the effects of dark matter and dark energy using the Branse-Dicke dilaton $\Phi$.
 To outline the basic properties of the MDG model of SSSS and to compare them with general relativistic results,
 in the present paper we use the relativistic equation of state (EOS) of neutron matter as an ideal Fermi neutron gas at zero temperature.
 We overcome the well-known difficulties of the physics of SSSS in the f(R) theories of gravity \cite{Felice10,Berti}
 applying novel highly nontrivial nonlinear boundary conditions, which depend on the global properties of the solution and on the EOS.
 We also introduce two pairs of new notions: cosmological-energy-pressure densities
 and dilaton-energy-pressure densities, as well as two new EOS for them: cosmological EOS (CEOS) and dilaton EOS (DEOS).
 Special attention is paid to the dilatonic sphere (in brief -- disphere) of SSSS,
 introduced in this paper for the first time. Using several realistic EOS for neutron star (NS):
 SLy, BSk19, BSk20 and BSk21, and current observational two-solar-masses-limit,
 we derive an estimate for scalar-field-mass $m_\Phi \sim 10^{-13} eV/c^2 \div 4\times 10^{-11} eV/c^2$.
 Thus, the present version of the paper reflects some of the recent developments of the topic.

\end{abstract}

\ccode{PACS Nos.: include PACS Nos.}

\section{Introduction}
The MDG model was proposed and studied in \cite{OHanlon72,Fiziev00a,Fiziev02,Fiziev13}.
Some of its applications and properties can also be found in \cite{Boisseu00, Fiziev00b,EFarese,Fiziev03}.
It describes a proper generalization of the Einstein general relativity (GR), being
based on the following action of the gravi-dilaton sector
\begin{equation}
{\cal A}_{g,\Phi}={\frac c {2\kappa}}\int d^4 x\sqrt{|g|}
 \bigl( \Phi R - 2 \Lambda U(\Phi) \bigr).
\label{A_MDG}
\end{equation}
Without any relation with astrophysics and cosmology,
MDG was studied for the first time by O'Hanlon, as early as in \cite{OHanlon72}.
His goal was to justify the Fujii idea about the "fifth force".
There the term "dilaton" for the Branse-Dicke field $\Phi$ \cite{BD} was introduced.

To some extent, MDG corresponds to the Branse-Dicke theory with the identically vanishing parameter $\omega$
and additional potential $U(\Phi)$.
This property is also used as the name of the MDG  model in some publications on the $f(R)$
theories of gravity, see \cite{Berti,Starobinsky80,Starobinsky07,Faraoni06,Frolov08,Felice10,Sotiriou10,Clifton12}
and a huge amount of references therein.
Since the presence of the potential $U(\Phi)$ yields a radically different physics in comparison
with the original Branse-Dicke theory, there is no physical reason to prescribe this name to MDG.

In general, the $f(R)$ theories are only {\em locally} equivalent to MDG, see \cite{Fiziev13}
where the critically important, from a physical point of view, class of withholding potentials
was introduced for the first time.
In the large literature on the $f(R)$ theories one is not able to find examples of $f(R)$-models
which own withholding property and are globally equivalent to MDG with the same physical properties.

In Eq. \eqref{A_MDG}, $\kappa=8\pi G_{N}/c^2$ is the Einstein constant, $G_N$ is the Newton gravitational constant,
$\Lambda$ is the cosmological constant, and $\Phi\in (0,\infty)$ is the dilaton field. The values $\Phi$ must be positive
since a change of the sign of $\Phi$ entails a change of the sign of the gravitational factor $G_N/\Phi$
and leads to antigravity instead of gravity.
Such a change is physically unacceptable. Besides, the value $\Phi=0$
must be excluded since it leads to an infinite gravitational factor and makes the Cauchy problem
in MDG not well posed \cite{EFarese}.
The value $\Phi=\infty$ turns off the gravity and is also physically unacceptable.

The Branse-Dicke field $\Phi$ is introduced to consider a variable gravitational factor $G(\Phi)=G_N/\Phi=G_N g(\Phi)$ instead of the gravitational constant $G_N$.
The cosmological potential $U(\Phi)$ is introduced to consider  a variable cosmological factor $\Lambda U(\Phi)$
instead of the cosmological constant $\Lambda$. In GR with cosmological constant $\Lambda$
we have $\Phi\equiv 1$, $g(\Phi)\equiv 1$, and $U(1)\equiv 1$.
Due to its specific physical meaning,
the field $\Phi$ has quite unusual properties.

The function $U(\Phi)$ defines the cosmological potential
which must be a positive {\em single valued} function of the dilaton field $\Phi$ by astrophysical reasons \footnote{The
astrophysical observations show that in the observable Universe the cosmological term in Eq. \eqref{A_MDG} has a unique negative value,
$\Lambda$ being a positive constant.}.
See \cite{Fiziev13} for all physical requirements on the cosmological potential $U(\Phi)$, which are
necessary for a sound MDG model.
There the class of {\em withholding} potentials was introduced.
These confine dynamically the values of the dilaton $\Phi$ in the physical domain.
It is hard to formulate such a property for the function $f(R)$ in a simple intuitive way.

In \cite{OHanlon72,Fiziev00a,Fiziev00b,Fiziev02,Boisseu00,Fiziev03,Fiziev13,EFarese}, one can find a comparison of MDG with observations, observational restrictions on the mass of the dilaton  $m_\Phi$, discussions of the MDG-cosmology, and some study of the structure of boson stars in MDG.
An important physical comment on possible relation of MDG with quantum gravity may be found in \cite{Fiziev02,Fiziev03}.

The f(R) gravity may have difficulties with the existence of stable star
configurations and the presence of a different type of singularities inside the stars, see for example
\cite{Berti,Frolov08,Felice10,Sotiriou10,Clifton12,Kobayashi08}.
There exist a large number of articles on stars which combine different f(R) or other models of gravity with
different EOS for star matter, see, for example, \cite{Nojiri08b,Babichev09,Babichev10,Astashenok14,Katsuragawa16}
and the references therein.

On the other hand, by representing $f(R)=R+\Delta f(R)$ with small perturbation $\Delta f(R)$
one obtains models of SSSS which deviate not very much
from the corresponding GR models \cite{Arapoglu11,Deliduman12}.

The main goal of the present article is to construct an example of a physically consistent family of SSSS
in MDG using the simplest withholding potential $U(\Phi)$ and simplest matter EOS (MEOS).

We explicitly show how the dilatonic field $\Phi$ changes the structure of the compact stars
and creates a specific disphere around them.

\section{Basic equations and boundary conditions for SSSS in MDG}
In units $G_N=c=1$  the field equations of MDG can be written in the form:
\begin{equation}
\Phi \hat{R}_\alpha^\beta +\widehat{\nabla_\alpha\nabla^\beta}\Phi+ 8\pi\hat{T}_\alpha^\beta = 0, \qquad
\Box\Phi+ \Lambda V^\prime(\Phi) = {\frac {8\pi} 3} T.
\label{DGE}
\end{equation}
Here ${T}_\alpha^\beta$ is the standard energy-momentum tensor of the matter, $\hat{X}_\alpha^\beta={X}_\alpha^\beta - {\frac 1 4}X\delta_\alpha^\beta$
is the traceless part of any tensor ${X}_\alpha^\beta$ in four dimensions,
$X=X_\alpha^\alpha$ is its trace, the relation $V^\prime(\Phi)={\frac 2 3}\Big(\Phi U^\prime(\Phi) -2U(\Phi)\Big)$ introduces the dilatonic potential $V(\Phi)$, and
the prime denotes differentiation with respect to the variable $\Phi$. See for conventions and notation \cite{Fiziev13}.

In the problems under consideration, the space-time-interval is
$ds^2=e^{\nu(r)}dt^2-e^{\lambda(r)}dr^2 - r^2 d\Omega^2 $ \cite{Landau},
where $r$ is the luminosity distance to the center of symmetry,
and $d\Omega^2$ describes the space-interval on the unit sphere.
Then, after some algebra one obtains  the following basic results
for a SSSS of the luminosity radius $r^*$.

In the inner domain   $r\in [0,r^*]$  the SSSS structure is determined by the system:
\begin{eqnarray}
{\frac {dm}{dr}}&=&4\pi r^2\epsilon_{eff}/\Phi, \label{DE:a}\\
{\frac {d\Phi}{dr}}&=&-4\pi r^2 p_{{}_\Phi}/\Delta,  \label{DE:b}\\
{\frac {dp_{{}_\Phi}}{dr}}&=&- {\frac{ p_{{}_\Phi}}{r\Delta}}\left(3r -7 m-{\frac 2 3}\Lambda r^3+4\pi r^3\epsilon_{eff}/\Phi\right)-{\frac{2}{r}}\epsilon_{{}_\Phi}, \label{DE:c}\\
{\frac {dp}{dr}}&=&- {\frac {p+\epsilon}{r}}\,{\frac{m+4\pi r^3 p_{eff}/\Phi}{\Delta-2\pi r^3 p_{{}_\Phi}/\Phi} },
\label{DE:d}
\end{eqnarray}
written in dimensionless variables, see the Appendix A.

The four unknown functions are $m(r)$, $\Phi(r)$, $p_{{}_\Phi}(r)$, and $p(r)$. In Eqs. \eqref{DE:a} -- \eqref{DE:d}
$\Delta=r-2 m-{\frac 1 3}\Lambda r^3$,
\begin{equation}
\epsilon_{eff}=\epsilon+\epsilon_{{}_\Lambda}+\epsilon_{{}_\Phi}, \quad p_{eff}=p+p_{{}_\Lambda}+p_{{}_\Phi}.
\label{ep_total}
\end{equation}
In addition, we obtain two novel EOS, which are specific for MDG:
\begin{eqnarray}
\epsilon_{{}_\Lambda}&=&- p_{{}_\Lambda}-{\frac \Lambda {12\pi}}\Phi \,;\label{NewEOS:a}\\
\epsilon_{{}_\Phi}&=&p-{\frac 1 3}\epsilon +
{\frac \Lambda {8\pi}}V^\prime(\Phi)+{\frac {p_{{}_\Phi}} 2}\,{\frac{m+4\pi r^3 p_{eff}/\Phi}{\Delta-2\pi r^3 p_{{}_\Phi}/\Phi}}\,; \label{NewEOS:b}\\
\epsilon &=& \epsilon(p). \label{NewEOS:c}
\end{eqnarray}
Equation \eqref{NewEOS:a} is the CEOS for the cosmological energy density
$\epsilon_{{}_\Lambda}$ and the cosmological pressure
$p_{{}_\Lambda}$:
\begin{equation}
\epsilon_{{}_\Lambda}={\frac \Lambda {8\pi}} \Big(U(\Phi)-\Phi\Big),\quad
p_{{}_\Lambda}=-{\frac \Lambda {8\pi}} \Big(U(\Phi)-{\frac 1 3}\Phi\Big).
\label{epLambda}
\end{equation}
These new quantities depend on the values of the dilaton $\Phi$ and on the values of the cosmological potential $U(\Phi)$.

Equation \eqref{NewEOS:b} follows from Eq. \eqref{DE:b} and presents the DEOS
for dilatonic energy density
$\epsilon_{{}_\Phi}={\frac 1 {8\pi r^2}}\left(\Delta/r\right)^{1/2}{\frac d {dr}}\left(r^2 \left(\Delta/r\right)^{1/2}{\frac d {dr}}\Phi\right)$
and  dilatonic pressure $p_{{}_\Phi}=-{\frac \Delta {4\pi r^2}}  {\frac d {dr}}\Phi$.
These new quantities depend on the gradient of gravitational factor with respect to the luminosity radius $r$.

Using the area $A=4\pi r^2$ of the surrounding sphere and
the true geometrical distance  $dl$ defined by the representation of the four-interval in the form
$ds^2=\alpha(l) dt^2+dl^2+A(l)/{4\pi}d\Omega^2 \Rightarrow dl= dr\big/ \sqrt{1- {\tfrac {2m} r} - {\tfrac 1 3}\Lambda r^2}$, we obtain much more compact expressions
\begin{equation}
\epsilon_{{}_\Phi}={\frac 1 {8\pi} }{\frac 1 A } {\frac d {dl} }\left(\! A {\frac {d\Phi} {dl}}\right), \quad p_{{}_\Phi}={\frac 1 {8\pi} }{\frac 1 A }{\frac {dA} {dl}}{\frac {d\Phi} {dl}}.
\label{epPhi}
\end{equation}

Equation \eqref{NewEOS:c} presents the usual MEOS of star matter, see, for example, \cite{Luciano} for a modern detailed survey.

Adopting the widespread assumption that the SSSS-center C ($\Rightarrow$ index "c") is at $r_c=0$ (i.e., when $A=0$), we obtain the boundary conditions
\begin{eqnarray}
m(0)=m_c=0,\quad \Phi(0)=\Phi_c,\quad p(0)=p_c,\nonumber\\
p_{{}_\Phi}(0)=p_{\Phi c}={\frac 2 3}\left({\frac {\epsilon(p_c)} 3}- p_c\right)-{\frac{\Lambda}{12\pi}}V^\prime(\Phi_c).
\label{center}
\end{eqnarray}
Requiring $m_c=0$, we ensure finiteness of pressure $p_c$ simultaneously for the Newton-, GR- and MDG-SSSS.
The condition on $p_{\Phi c}\,(=-{\frac 2 3}\epsilon_{\Phi c})$  ensures its finiteness,
being a specific MDG-centre-values-relation: $F_\Phi(p_{\Phi c},p_c,\Phi_c)=0$.

The SSSS-edge ($\Rightarrow$ index "*")  is defined by the condition $p^*=p(r^*;p_c,\Phi_c)=0$ (and typically $\epsilon^* =0$). Then
\begin{equation}
m^*=m(r^*;p_c,\Phi_c),\quad \Phi^*=\Phi(r^*;p_c,\Phi_c),\quad p_{{}_\Phi}^*=p_{{}_\Phi}(r^*;p_c,\Phi_c).
\label{edge}
\end{equation}
The luminosity radius of a compact star may vary to some extent for different physically sound MEOS.
In general, the safe limits seem to be $r^*\in [5, 20]$ km.

Outside the star $p\equiv 0$ and $\epsilon\equiv 0$, and we have a dilaton-sphere,
in brief -- {\em a disphere}. Its structure is determined
by the shortened system \eqref{DE:a}--\eqref{DE:d}: Eq. \eqref{DE:d} has to be omitted.
For the exterior domain $r\in [r^*,r_{{}_{\!U}}]$ we use Eqs. \eqref{edge} as left boundary conditions.
The right boundary is defined by the cosmological horizon with unknown position $r_{{}_{\!U}}$
where the de Sitter vacuum is reached. Thus, we obtain a new {\em nonlocal} system of equations
\begin{equation}
\Delta(r_{{}_{\!U}}; p_c,\Phi_c)=0, \quad \Phi(r_{{}_{\!U}}; p_c,\Phi_c)=1
\label{nonlocal}
\end{equation}
which relates the values of $r$ and $\Phi$ at different space-time points
and is defined by the global structure of the space-time with a single compact SSSS in it.

The elimination of the unknown quantity $r_{{}_{\!U}}$ from Eqs. \eqref{nonlocal}
leads to the second MDG-centre-values-relation:   $F_\Lambda(p_c,\Phi_c)=0$.

In Fig. \ref{Fig1} (Left), we present for the first time a numerical solution of this novel nonlocal physical problem.
This highly nontrivial solution strongly depends also on the MEOS and on the star interior structure
(compare the result of the present paper with the results of the subsequent ones \cite{Fiziev15a,Fiziev15b,Fiziev17}.).

The two MEOS-dependent-relations
\begin{equation}
F_\Phi(p_{\Phi c},p_c,\Phi_c)=0,\,\,\,F_\Lambda(p_c,\Phi_c)=0,
\label{F}
\end{equation}
show that in MDG, as well as in the Newton gravity and GR, we have a one-parameter-family of SSSS.
This very important specific property of the MDG model of SSSS
is not typical of other extended theories of gravity.

In dimension-full variables the observable  value $\Lambda \sim 10^{-44}$ $km^{-2}$ is very small.
As a result, the luminosity radius of the Universe $r_{{}_{\!U}} \sim 1/\sqrt{\Lambda}\sim 10^{22}\,\text{km}$
is very large in comparison with the typical luminosity radius of the compact star $r^*\sim 10$ km.
This circumstance generates hard numerical problems and a need for special computational methods.

Further on, we use the cosmological potential
$U(\Phi)=\Phi^2+{\frac{3}{16\, \mathfrak{p}^2}}\left(\Phi-1/ \Phi\right)^2$. Hence,
$V^\prime(\Phi)={\frac 1 {2 \mathfrak{p}^2}}\left(1-1/\Phi^2\right)$.
For useful comments and a more general form of the admissible cosmological potentials $U(\Phi)$ see \cite{Fiziev02,Fiziev13}.
The parameter $\mathfrak{p}=\sqrt{\Lambda}\,\hbar/c\,m_\Phi$ is the dimensionless
Compton length (measured in cosmological units) of the dilaton $\Phi$.
MDG is consistent with observation if $\mathfrak{p}\lesssim 10^{-30}$ (i.e., if $m_\Phi> 10^{-3}\, \text{eV}/c^2)$ \cite{Fiziev00a,Fiziev02}.

At present, the experimental determination of the mass of the dilaton is the most significant physical problem
of MDG. For more complicated cosmological potentials it is possible to have several values
of this mass which correspond to different local  minima of the dilaton potential $V(\Phi)$ \cite{Fiziev13}.
If one considers as a dilaton the only currently known fundamental scalar particle,
the Higgs boson with mass $m_H \approx 125\, GeV/c^2$, then $\mathfrak{p} \approx 1.8\times 10^{-43}$.

In our numerical calculations in Section 3,
we  use a maybe non-realistic large value $\mathfrak{p} =  10^{-21}$ (the Compton length of the dilaton
is $\sim 9$ km being comparable with a typical value of the NS star luminosity radius $r^*$)
just for a more transparent graphical representation of our qualitatively new results.

\section{The results for the simplest EOS for neutron matter}

We do not present here in detail a model of a neutron star with realistic MEOS.
The most idealized relativistic MEOS of neutron matter is the well-known one
used in the Tolman-Oppenheimer-Volkov (TOV) model  \cite{TOV}. In dimensionless variables it reads
\begin{equation}
p={\tfrac 1 {12\pi}} \left( \sinh t -8\sinh(t/2)+3t\right), \quad \epsilon= {\tfrac 1 {4\pi}} \left( \sinh t -t\right).
\label{IFNG0T}
\end{equation}
It describes the ideal Fermi neutron gas at zero temperature and is free from
the difficulties related with unknown properties of neutron matter.
The possible values of $\epsilon>0$ and $p>0$ are not a priori  constrained from the above.

The analytic form of MEOS \eqref{IFNG0T} facilitates our study of the new pure-MDG-effects in SSSS, shown in Figs. \ref{Fig1}--\ref{Fig6}.
For units see Appendix A.
When applicable, we use different line styles for star's interior and for star's exterior, as well as for GR-results and for MDG-results.
\begin{figure}[!ht]
\begin{center}
\begin{minipage}{12.cm}
\vskip -.truecm
\hskip .5truecm
\includegraphics[width=6.1truecm]{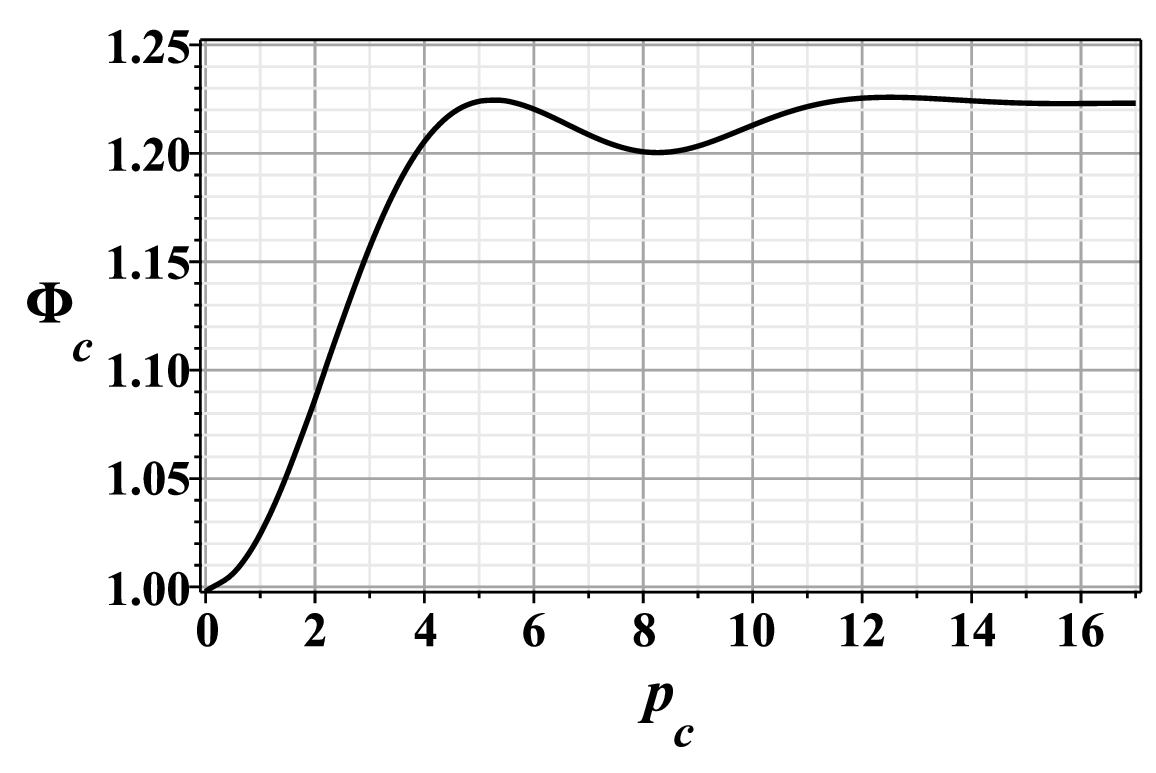}
\hfill
\vskip -4.truecm
\hskip 7.5truecm
\includegraphics[width=3.9truecm]{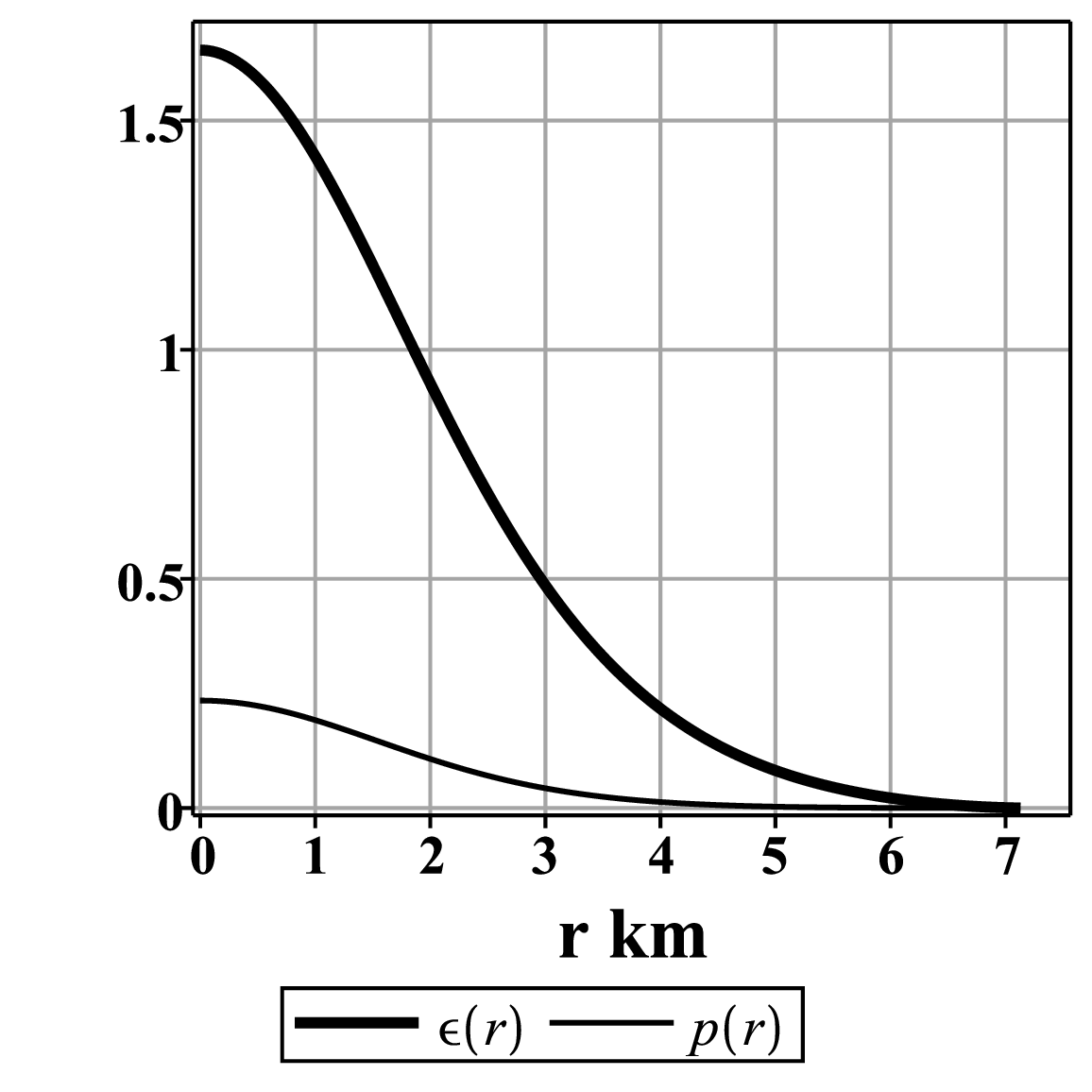}
\vskip .5truecm
\caption{\small \hskip .truecm Left: The specific MDG-curve $F_\Lambda(p_c,\Phi_c)=0$.}
 \small Right: The MDG-SSSS interior for MEOS \eqref{IFNG0T}: dimensionless energy density $\epsilon(r)$ and pressure $p(r)$.
\label{Fig1}
\end{minipage}
\end{center}
\end{figure}
\begin{figure}[!ht]
\begin{minipage}{12.cm}
\vskip -1.truecm
\hskip .truecm
\includegraphics[width=3.8truecm]{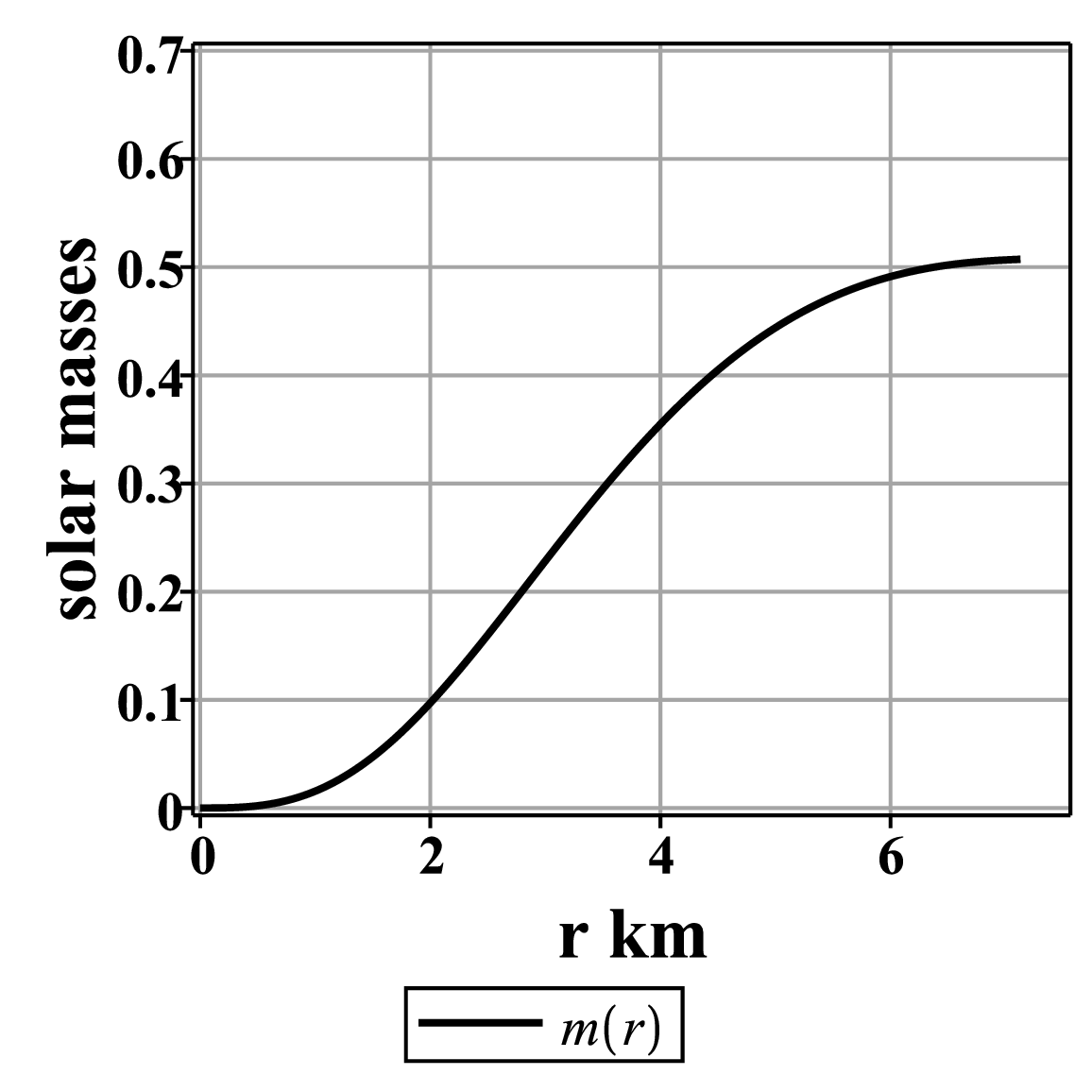}
\hfill
\vskip -3.8truecm
\hskip 5.5truecm
\includegraphics[width=6.5truecm]{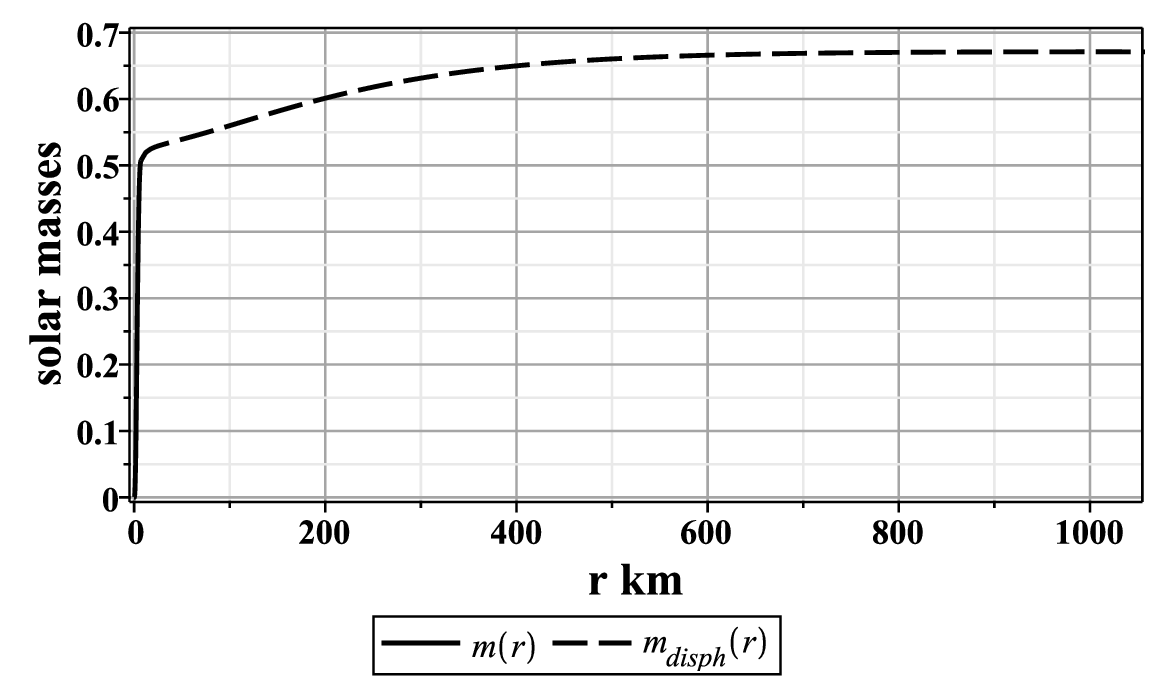}
\vskip .truecm
\caption{\small Left: The MDG-SSSS mass distribution $m(r)$ in the star interior for MEOS \eqref{IFNG0T}.
 Right: The SSSS-disphere-mass-dependence on $r$}
\label{Fig2}
\end{minipage}
\end{figure}
\begin{figure}[!ht]
\begin{minipage}{12.cm}
\vskip .5truecm
\hskip -.truecm
\includegraphics[width=5.7truecm]{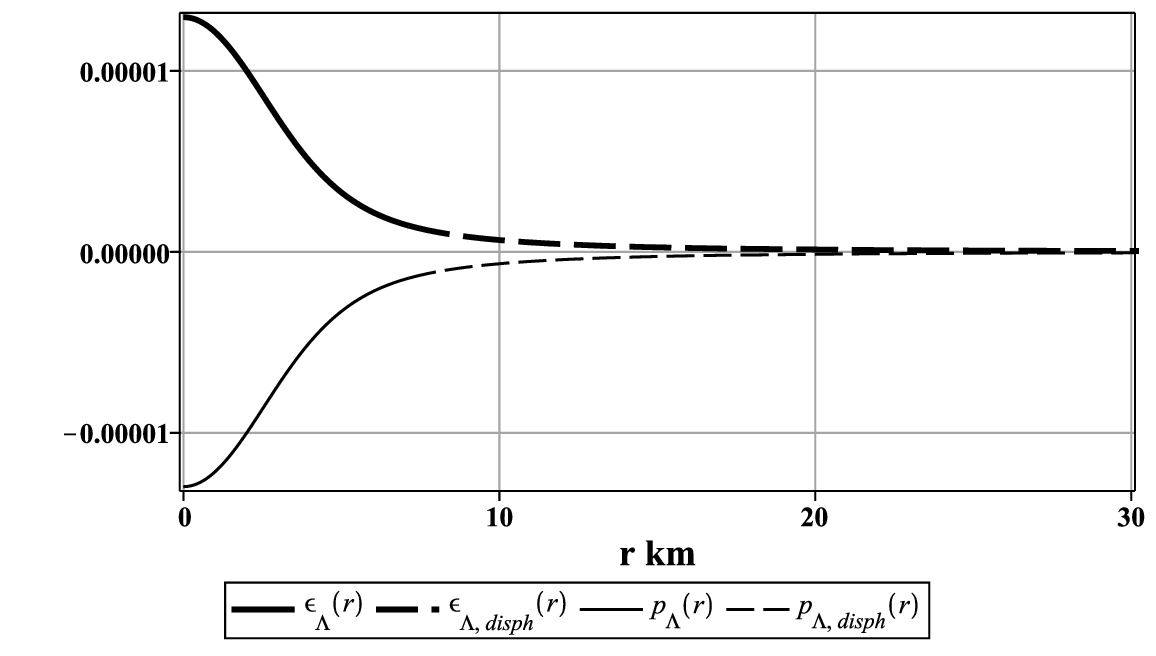}
\hfill
\vskip -3.2truecm
\hskip 6.5truecm
\includegraphics[width=5.truecm]{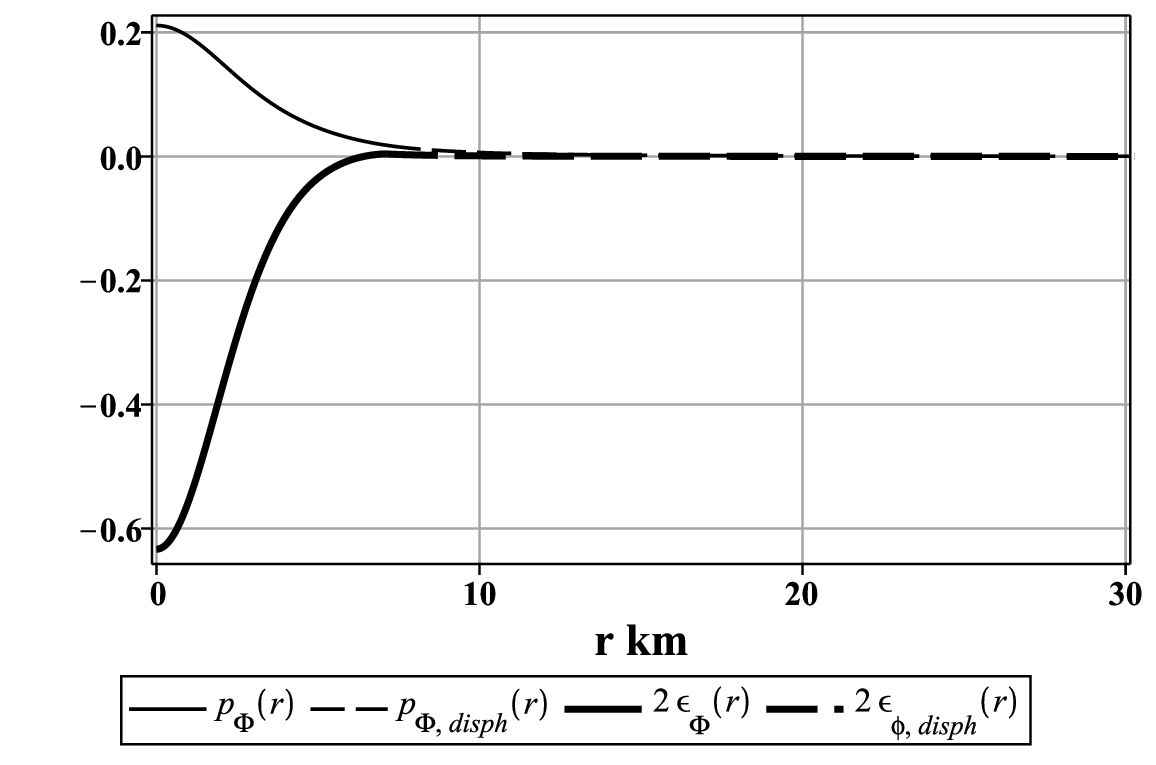}
\vskip .truecm
\caption{\small Left: The  dimensionless functions $\epsilon_\Lambda(r)$ and $p_\Lambda(r)$ for CEOS \eqref{NewEOS:b}.
 Right: The  dimensionless functions $2\epsilon_\Phi(r)$ and $p_\Phi(r)$ for DEOS \eqref{NewEOS:c} }
\label{Fig3}
\end{minipage}
\end{figure}
\begin{figure}[!ht]
\begin{minipage}{12.cm}
\vskip -.5truecm
\hskip -.truecm
\includegraphics[width=5.5truecm]{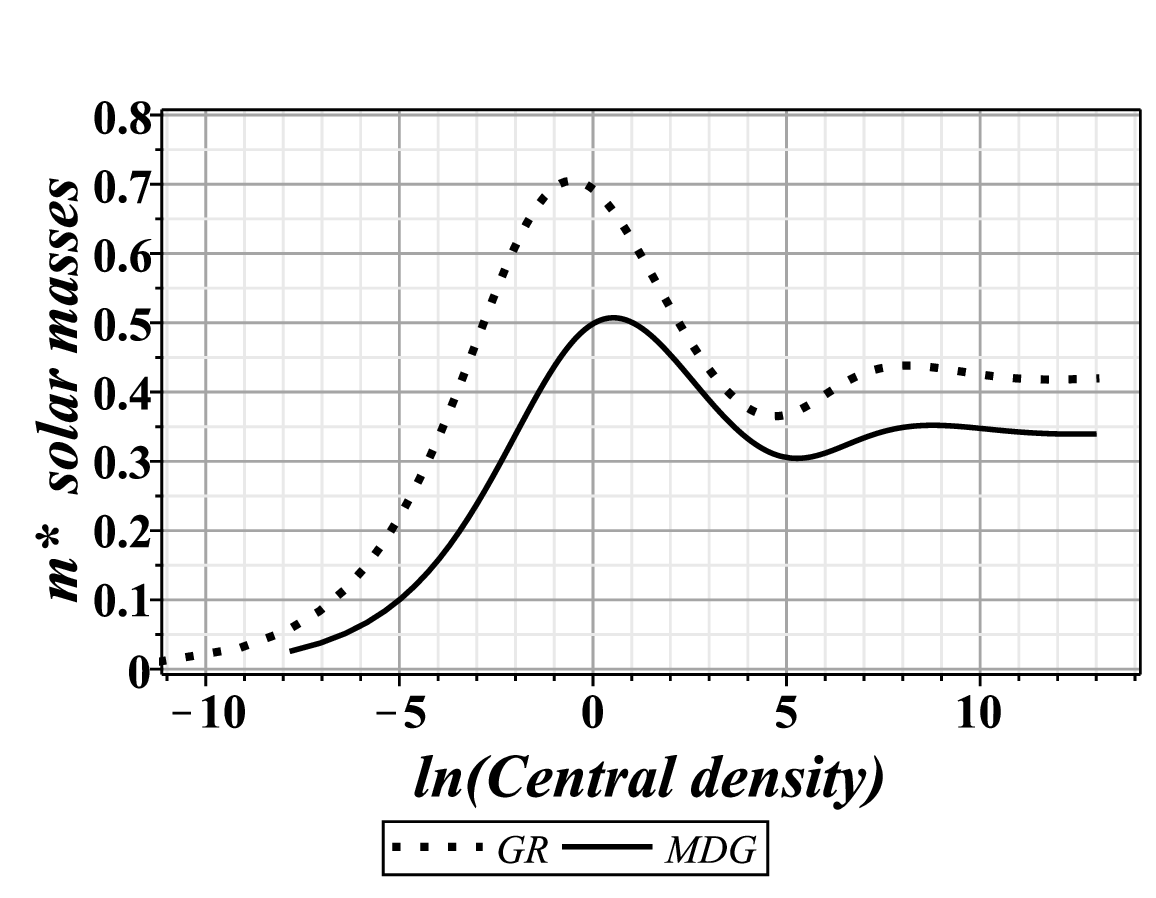}
\hfill
\vskip -4.3truecm
\hskip 6.5truecm
\includegraphics[width=5.5truecm]{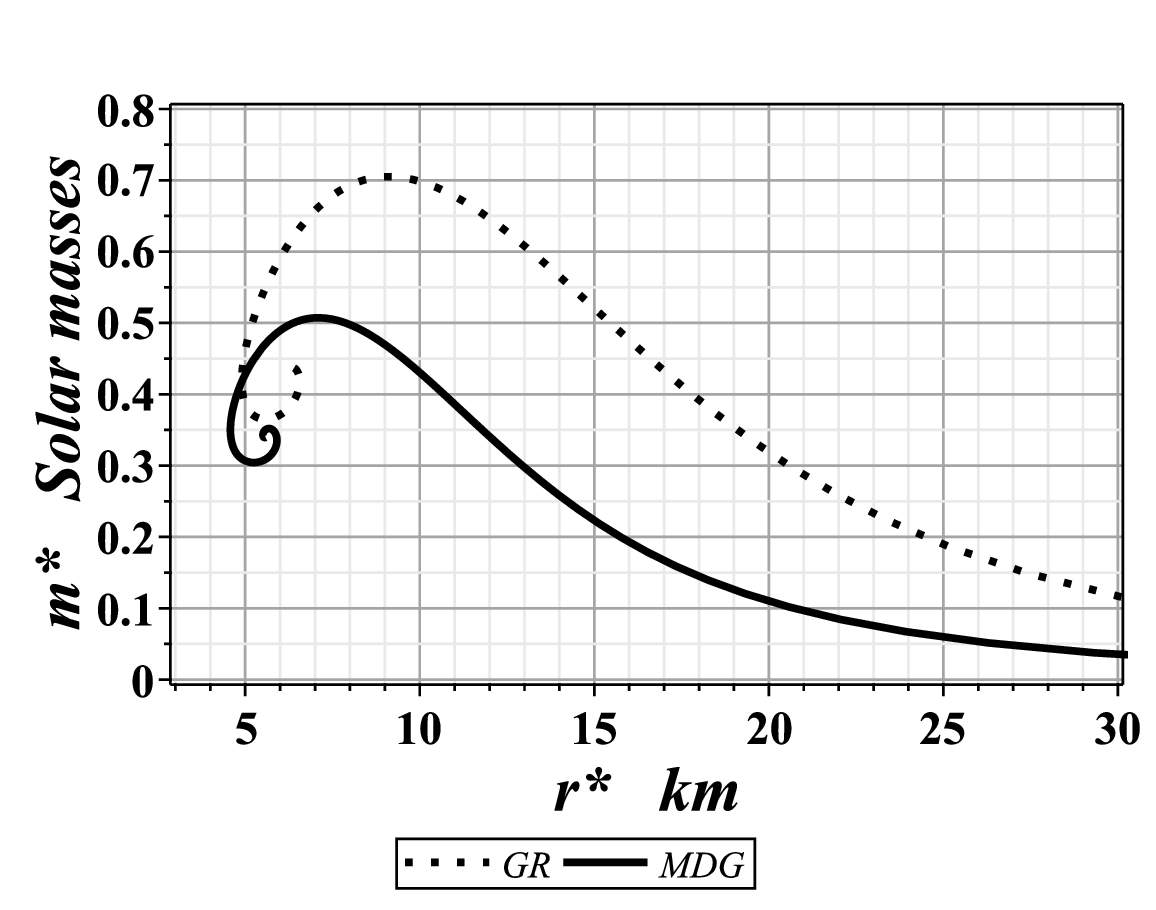}
\vskip .truecm
\caption{\small Left: The m* -- $\epsilon_c$ dependencies.
 Right:  Mass - radius relations for IFNG0T in
\hskip 0.truecm GR ($m_{max}^*\approx .7051\,m_{\odot}$, $r_{max}^*\approx 9.209$ km) and
\hskip 0.truecm in MDG ($m_{max}^*\approx .5073\,m_{\odot}$, $r_{max}^*\approx 7.092$ km)}
\label{Fig4}
\end{minipage}
\end{figure}
\begin{figure}[!ht]
\begin{minipage}{12.cm}
\vskip -.truecm
\hskip -.truecm
\includegraphics[width=5.2truecm]{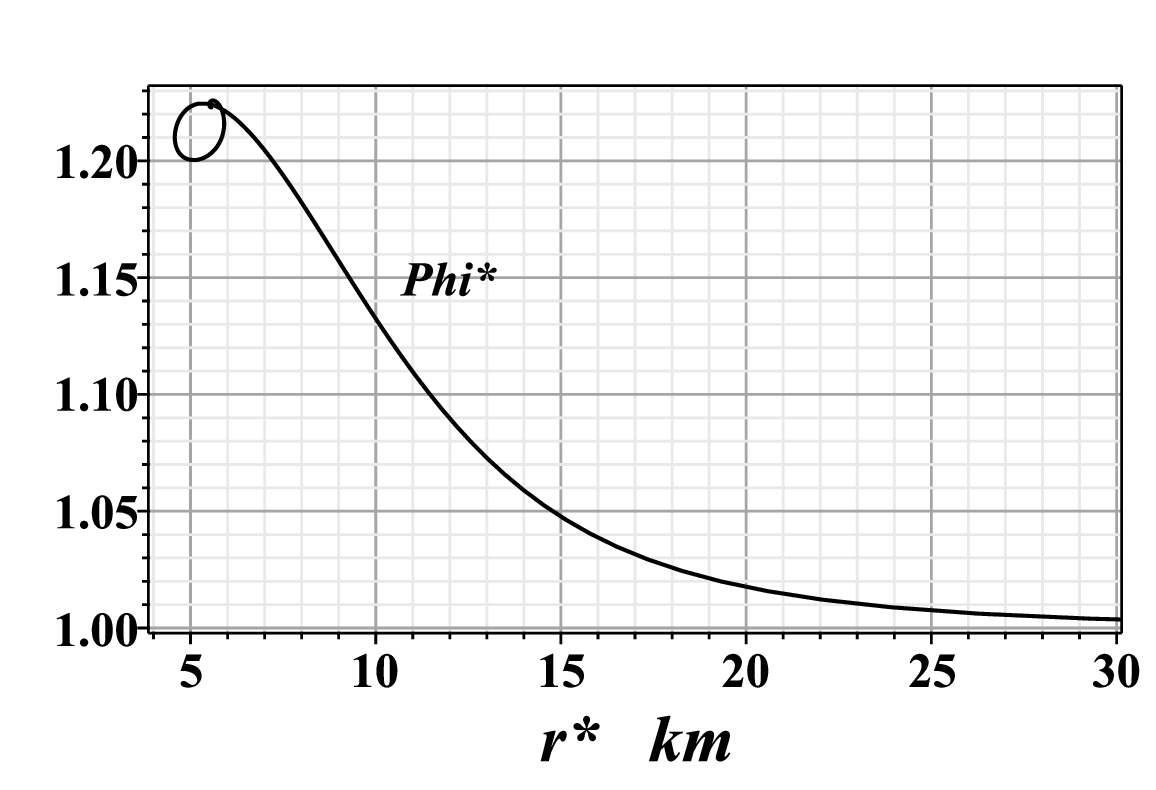}
\hfill
\vskip -3.3truecm
\hskip 6.truecm
\includegraphics[width=5.2truecm]{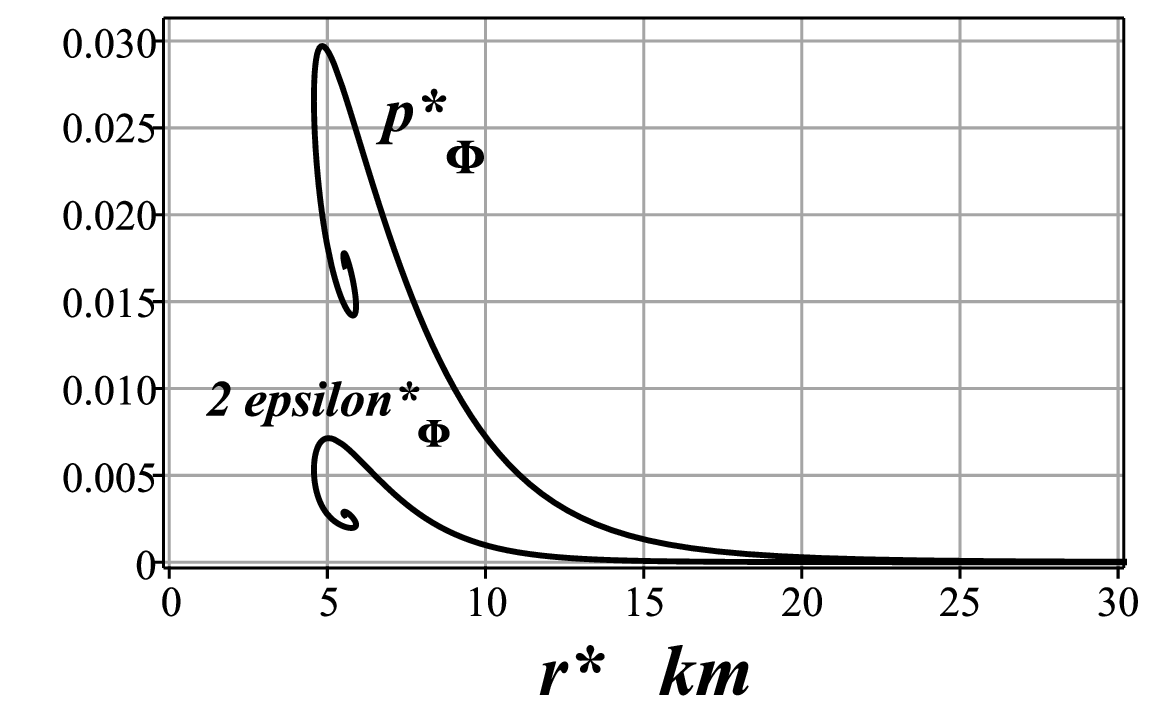}
\vskip .truecm
\caption{\small Left: The specific MDG-dependence $\Phi^*(r*)$.
 Right: The specific MDG-dependences $2 \epsilon_\Phi^*(r*)$, $p_\Phi^*(r*)$ (dimensionless).
Maximal values are reached for smaller $r*$ than in Fig. \ref{Fig4} (Right)}
\label{Fig5}
\end{minipage}
\end{figure}
\begin{figure}[!ht]
\begin{minipage}{12.cm}
\vskip -2.1truecm
\hskip -.truecm
\includegraphics[width=5.8truecm]{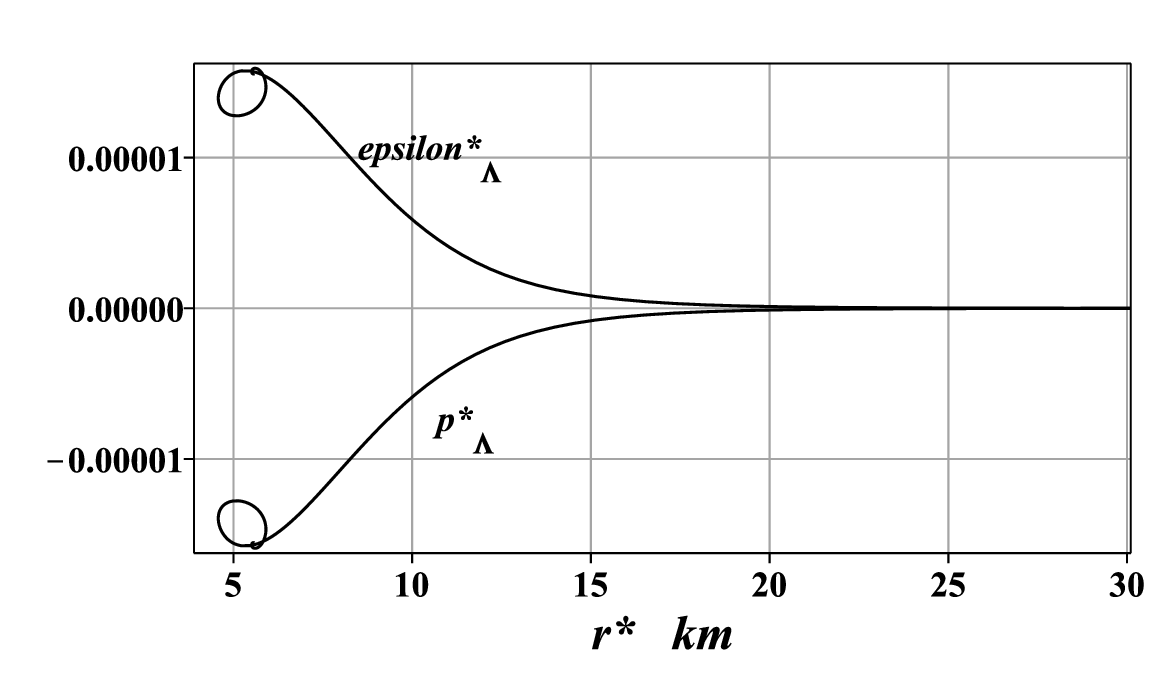}
\hfill
\vskip -3.2truecm
\hskip 6.truecm
\includegraphics[width=5.2truecm]{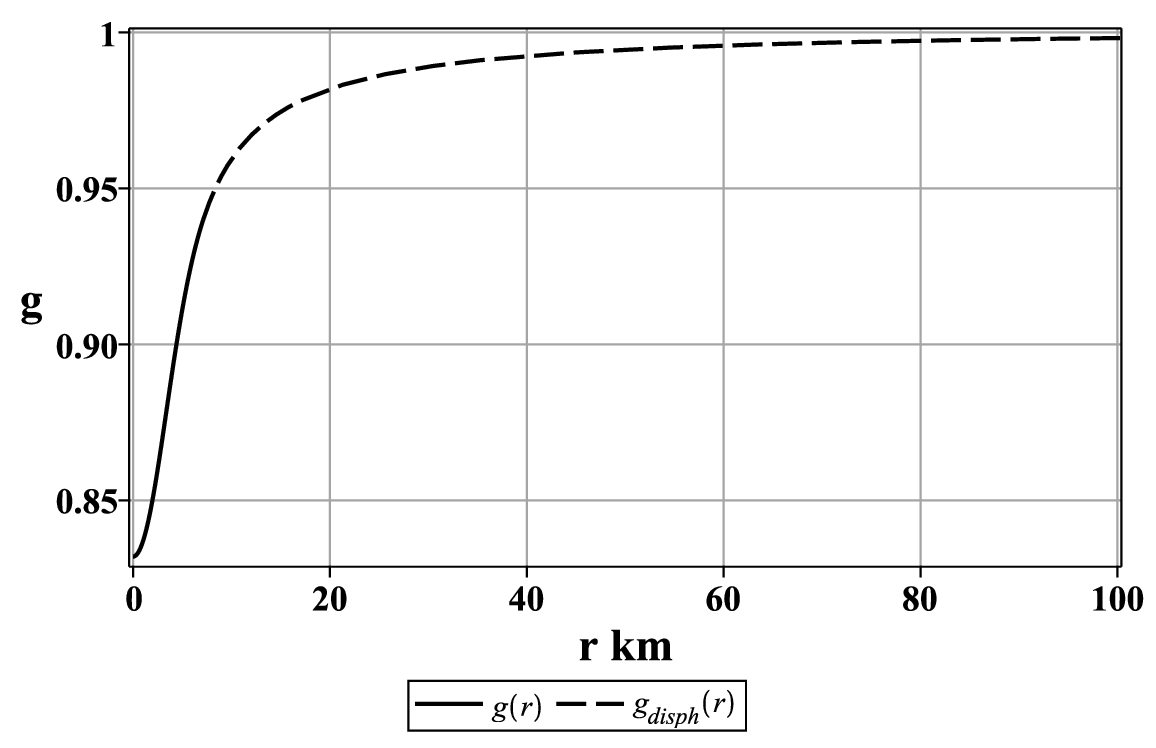}
\vskip .truecm
\caption{\small Left: The specific MDG-dependences  $\epsilon_\Lambda^*(r*)$ and $p_\Lambda^*(r*)$ (dimensionless) in accord with CEOS.
 Right: The specific variations of the  dimensionless gravitational factor $g(r)=1/\Phi(r)$ for MDG-SSSS}
\label{Fig6}
\end{minipage}
\end{figure}

\newpage

\section{Comments of the numerical results}

\subsection{Comments of the numerical results for the MEOS \eqref{IFNG0T}}
The interior structure of MDG-SSSS is shown in Fig.\ref{Fig2} (Left).
It differs from the structure of GR-SSSS only quantitatively.

Figures \ref{Fig3} (Left) and \ref{Fig6} (Left) confirm our expectation that the $\Lambda$-term in action \eqref{A_MDG}
yields variable positive $\epsilon_\Lambda$ and variable negative pressure $p_\Lambda$ inside, as well as outside MDG-SSSS,
as it should be for description of dark energy effects.
The changes of  $\epsilon_\Lambda(r)$ and $p_\Lambda(r)$ are slightly asymmetric in accord with CEOS \eqref{NewEOS:a}.

As seen in  Fig. \ref{Fig4} (Right), MDG-SSSS with MEOS \eqref{IFNG0T} are lighter and more compact than GR-SSSS.

Precisely as in GR, MDG-SSSS may be stable only until maximal mass is reached, see Figs. \ref{Fig4} (Left) and \ref{Fig4} (Right).
As well as in GR, the full investigation of the stability problem of MDG-SSSS requires a study of their oscillations.

The similarity of Fig.\ref{Fig1} and  Fig.\ref{Fig4} (Left)
shows that they both follow the behaviour of the mass $m^*(r^*)$ -- Fig. \ref{Fig4} (Right).

One of the most interesting novel physical results is shown in Fig. \ref{Fig2} (Right):
The mass of the disphere $m_{disp}(r)$ outside MDG-SSSS as a function of the
luminosity radius $r\in[r^*, r_{{}_U}]$. As expected,
it exponentially goes to a constant. The total mass of the object exceeds the mass of the
very star. In our illustrative example the MDG-SSSS mass $ m^* \approx  .5073\, M_\odot$
is close to the extremal one.  We found the total mass of the disphere $m_{disphere} \approx .1638\, M_\odot$,
i.e. $\approx 32$ per cent of the mass of the star.
In the case under consideration, the total mass of the object
$m_{total} \approx 0.6710\, M_\odot$ is reached with good precision at the distance
of about several hundred of star radii from the center. The total mass is quite close to the mass of GR-SSSS.

\begin{figure}[!ht]
\begin{minipage}{13.5cm}
\vskip 4.truecm
\hskip -.2truecm
\includegraphics[width=5.7truecm]{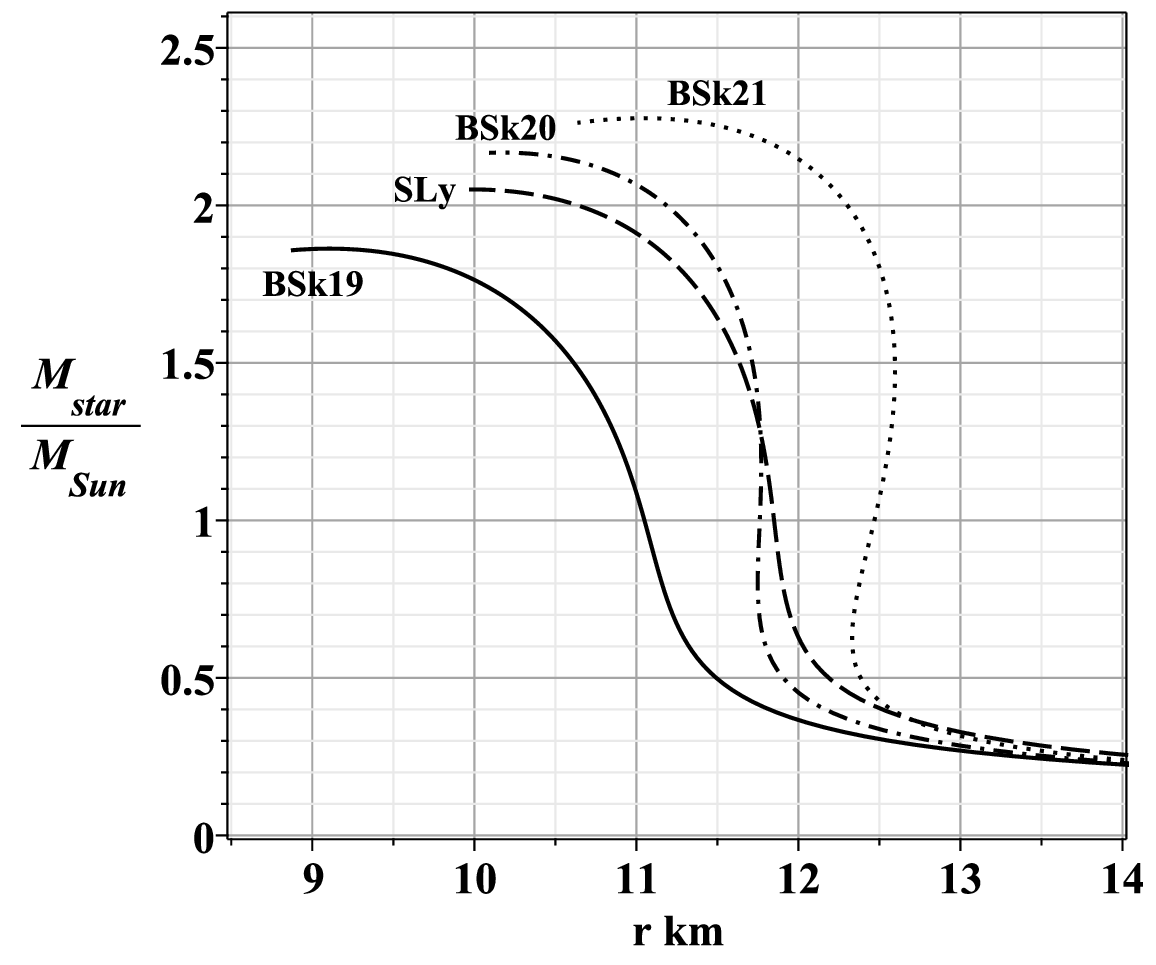}
\hfill
\vskip -4.8truecm
\hskip 5.7truecm
\includegraphics[width=7.4truecm]{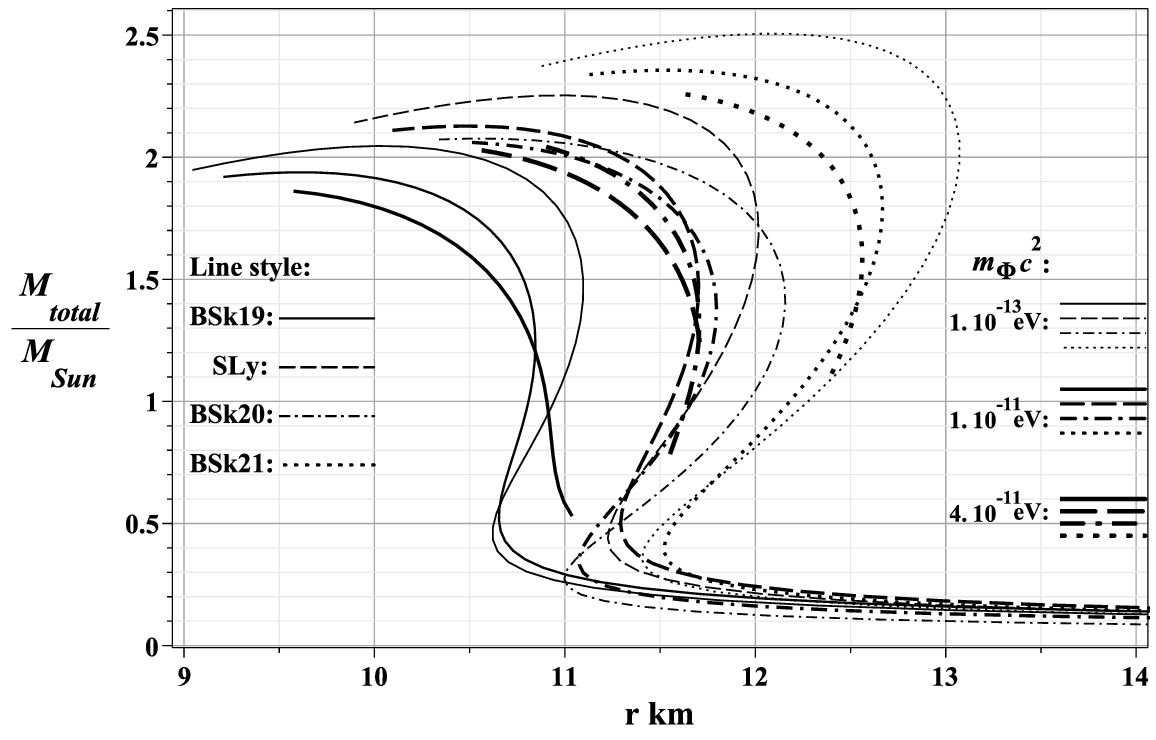}
\vskip 3.truecm
\caption{\small Left: The specific GR-mass-radius-dependencies in accord with MEOS BSk19, SLy, BSk20 and BSk21.
 Right: The specific MDG-mass-radius-dependencies in accord with MEOS BSk19, SLy, BSk20 and BSk21 and for different dilaton masses $m_\Phi$.}
\label{Fig7}
\end{minipage}
\end{figure}

\subsection{Numerical results for realistic MEOS SLy, BSk19, BSk20 and BSk21 and restrictions on the mass of dilaton
from two-solar-masses limit for NS}

The comparison of the above results with the results of articles on MDG-NS with realistic MEOS
\cite{Fiziev15a,Fiziev15b,Fiziev17} shows an interesting new phenomenon.
In the last cases the maximal masses $m^*$, as well maximal total masses $m_{tot}$
are larger than the maximal masses of NS in GR with the same MEOS and depend on dilaton mass $m_\Phi$, see Fig.\ref{Fig7}.

In contrast, the maximal GR-mass for MEOS \eqref{IFNG0T} is larger than the maximal MDG-mass.
A curious problem is to find a realistic MEOS (maybe a soft one) with the same maximal mass of NS in both GR and MDG.

Another interesting physical possibility is to use the MDG-model of NS to increase the maximal total mass of soft MEOS
in a way that will make it possible to reach the famous two-solar-masses limit for
PSRs J1614-22301: $(1.97 \pm 0.04)M\odot$ and J0348+0432  $(2.01 \pm 0.04)M\odot$  \cite{Lattimer09,Demorest10,Antoniadis13,Lattimer16}.
Then, the rejected at present types of soft MEOS, including the MEOS for quark and hybrid stars \cite{Ozel06}
may turn out to be compatible with observations due to the use of MDG.

Considering in detail the right Fig.\ref{Fig7} which shows mass-radius relations of MDG-NS
with realistic MEOS BSk19, SLy, BSk20 and BSk21 (see for more detail \cite{Fiziev15b}),
we conclude that the two-solar-masses limit allows MDG-NS with:
i) MEOS BSk19 and $m_\Phi \sim 10^{-11} eV/c^2$;
ii) MEOS SLy and $m_\Phi \sim 4\times 10^{-13} eV/c^2$; and
iii) MEOS BSk20 and $m_\Phi \in (10^{-11}, 4\times 10^{-11}) eV/c^2$;
but excludes  MDG-NS with MEOS BSk21 under the hypothesis that NS of larger masses does not exist in Nature.
In contrast, for GR-NS the two-solar-masses limit allows only NS with MEOS SLy, see our left Fig.\ref{Fig7} and \cite{Chamel11,Fantina12,Potekhin13}.

The above novel estimate: $m_\Phi \sim 4\times 10^{-13}\div 10^{-11} eV/c^2$ obtained from the two-solar-masses limit for NS with realistic MEOS,
is about $8 \div 10$ orders of magnitude smaller than the earlier estimate: $m_\Phi \gtrsim 10^{-3} eV$,
obtained from modern experiments on the Earth surface,
see the references in \cite{Fiziev00a}, as well as the latest data in \cite{Hata09,Murata15,Hees17}.
In addition, in \cite{Starobinsky07,Starobinsky11,Starobinsky12} it was shown that from inflationary cosmology one obtains huge mass of dilaton:
$m_\Phi\sim 1.5\times 10^{-5} M_{pl}\approx 3.65\times 10^{22} eV/c^2$, $M_{pl}\approx 2.435\times 10^{18} GeV/c^2$ being the reduced Planck mass. The recent observational results of Planck mission \cite{Planck14} yield slightly lower $m_\Phi\sim 1.3\times 10^{-5} M_{pl}\approx 3.17\times 10^{22} eV/c^2$ \footnote{The author is thankful to the unknown referee for this remark}.

A natural way to avoid the obvious strong tension between these very different estimates in the framework of MDG seems to be the assumption that in Nature we have a withholding dilatonic potential $V(\Phi)$ with several minima \cite{Fiziev13}
that show up at different scales as different values of the dilaton mass $m_\Phi$.

\section{Concluding remarks}

The above consideration shows that the MDG enriches essentially the analysis of the realistic MEOS
and the criteria for their acceptance or exclusion. Much more work has to be done in this direction.

It will be quite interesting to work out a model of moving and rotating stars in MDG.
This is a much more complicated issue than the same problem in GR
because of the nonlinear boundary condition with the moving boundary
which must be consistent with the global structure of the MDG Universe.
In this case, one can expect not only a nonspherically symmetric configuration
but also the appearance of different centers of the star and its disphere,
or even detachment of parts of the disphere. Such a nonstatic model may
support strongly the possible interpretation of the dilaton field $\Phi$ as dark matter.
For similar effects, observed at scales of galactic clusters, see \cite{Clowe}.

Another problem of current importance is to find solution of the problem of binary-NS in MDG and the corresponding radiation of gravitational waves
during the binary merger. The presence of the disphere may also have an interesting additional contribution during such processes.

If the mass $m_\Phi$ around NS is small enough to yield a disphere of a large effective radius $\sim\hbar/(m_\Phi c)$,
then the disphere around NS can be observed via the Shapiro delay by using the method of \cite{Demorest10}.

The basic property of the MDG dilaton $\Phi$  is that it does not interact directly with matter
\cite{Fiziev00a,Fiziev02,Fiziev03,Boisseu00,Fiziev13}. Such interaction is possible only in quantum field
theory as a perturbative effect of the second order with respect to the small gravitational constant $G_N$.
Hence, the corresponding cross-sections will be extremely small
in accordance with the recent observational data,
see, for example, \cite{LUX} and the references therein.

All of the above results will remain valid also in the $f(R)$ theories of gravity,
which correspond to withholding MDG-potentials \cite{Fiziev13}.
Unfortunately, one cannot find such $f(R)$ theories of gravity in the large existing literature.

As a result of the above statements and the previous work on MDG,
a novel basic physical conjecture arises: to look simultaneously for realistic MEOS
and for withholding cosmological potentials which are able
to describe a variety of cosmological, astrophysical, star, planet and laboratory phenomena at different scales.

\vskip .truecm

\appendix

\section{Used units}

In full physical dimensions the system \eqref{DE:a}--\eqref{DE:d}  reads
\begin{eqnarray}
{\frac {dm}{dr}}&=&4\pi r^2c^{-2}\epsilon_{eff}/\Phi, \label{DEdim:a}\\\\
{\frac {d\Phi}{dr}}&=&-{\tfrac {G_N} {c^4}}4\pi r^2 p_{{}_\Phi}/\Delta,  \label{DEdim:b}\\
{\frac {dp_{{}_\Phi}}{dr}}&=&- {\frac{ p_{{}_\Phi}}{r\Delta}}\left(3r -7 {\tfrac {G_N} {c^2}}m-{\frac 2 3}\Lambda r^3+{\tfrac {G_N} {c^4}}4\pi r^3\epsilon_{eff}/\Phi\right)-{\frac{2}{r}}\epsilon_{{}_\Phi}, \label{DEdim:c}\\
{\frac {dp}{dr}}&=&- {\tfrac {G_N} {c^2}}{\frac {p+\epsilon}{r}}\,{\frac{m+4\pi r^3 c^{-2}p_{eff}/\Phi}{\Delta-{\tfrac {G_N} {c^4}}2\pi r^3 p_{{}_\Phi}/\Phi}},\label{DEdim:d}\\
\Delta &=& r - {\tfrac {G_N} {c^2}} 2m - {\tfrac 1 3}\Lambda r^3. \label{DEdim:e}
\end{eqnarray}

The dimension-full form of \eqref{IFNG0T} reads \cite{TOV}
\begin{equation}
p={\tfrac 1 3}K \left( \sinh t -8\sinh(t/2)+3t\right), \quad \epsilon= K \left( \sinh t -t\right),
\label{IFNG0Tdim}
\end{equation}
with $K =  m_n c^5 /(32\pi^2\hbar^{3})$, $m_n\approx 1.675\times 10^{-24} g$ being the mass of the neutron.

We obtain the dimensionless form of Eqs. \eqref{DE:a}--\eqref{DE:d} and \eqref{IFNG0T} making
transition to dimensionless variables $r,m,\epsilon,p,\epsilon_\Phi,p_\Phi,\epsilon_\Lambda,p_\Lambda,\Lambda$ and $K=1/4\pi$ \footnote{
The convenient, but otherwise arbitrary choice $K=1/4\pi$ fixes the units\cite{TOV}.} according to
\begin{equation}
r\rightarrow R_0 r,\,\,\, m\rightarrow M_0 m,\,\,\,\epsilon_{...} \rightarrow \epsilon_0 \epsilon_{...},
\,\,\,p_{...}\rightarrow \epsilon_0 p_{...},\,\,\,\Lambda\rightarrow \Lambda_0 \Lambda
\label{dimless}
\end{equation}
without changing the notation of variables. In Eqs. \eqref{dimless}
\begin{eqnarray}
R_0&=&{\tfrac {G_N} {c^2}}\left(\tfrac {8\pi}{m_n}\right)^2 M_{P}^3\approx 13.676 km,\,\,\,M_0=\left(\tfrac {8\pi}{m_n}\right)^2 M_{P}^3\approx 9.264 M_{\odot}, \nonumber\\
\epsilon_0&=&m_n c^5 /(8\pi\hbar^{3})\approx 6.47\times 10^{36}g\, cm^{-1} s^{-2},\nonumber\\
\Lambda_0 &=& R_0^{-2}\approx 5.35\times 10^{-3} km^{-2}.
\label{dimfull}
\end{eqnarray}
Here $M_P=\sqrt{\hbar c/8\pi G_n}\approx 4.341\times 10^{-6} g$ is the Planck mass.
Besides, we obtain the scale factor $\rho_0= m_n c^3 /(8\pi\hbar^{3})\approx 6.72\times 10^{15}g/cm^3$
for transition to dimensionless mass density $\rho \rightarrow \rho_0 \rho$.

The dimension-full physical values of the quantities in vertical axes in Figs. 1-Right, 3, 5 and 6-Left
can be obtained using the corresponding factors from Eq. \eqref{dimfull}.

\section*{Acknowledgments}

The author is deeply indebted to the Directorate of the
Laboratory of Theoretical Physics, JINR, Dubna, for the
good working conditions. This
research was supported in part by the Foundation for
Theoretical and Computational Physics and Astrophysics
and Grants of the Bulgarian Nuclear Regulatory
Agency for 2014, 2015, 2016, 2017.

The author is deeply thankful to the unknown referee for the useful suggestions,
for recommendations to include in the text Section 4.2 and  Appendix A,
for some corrections in the references, as well as for throwing author's attention to \cite{Boisseu00,Starobinsky12}.

The author is also grateful to Kalin Marinov for the useful discussion
of the NS models in the MGD during his current PhD student studies.

\end{document}